\documentclass[%
preprint,
 amsmath,amssymb,
 aps,
]{revtex4-2}
\usepackage{tikz}
\usetikzlibrary{decorations.pathmorphing,decorations.markings,arrows.meta}

\usepackage{graphicx}
\usepackage{dcolumn}
\usepackage{bm}

\usepackage{physics}
\usepackage{subcaption}

\begin{document}

\title{Quantum Geometry and Pair Breaking in Intra- and Intervalley Superconductivity of Rhombohedral Graphene}
\author{Louis Primeau}
\affiliation{Theoretical Division, Los Alamos National Lab}
\affiliation{Department of Physics and Astronomy, University of Tennessee, Knoxville, Tennessee 37996, USA}
\author{Yang Zhang}
\affiliation{Department of Physics and Astronomy, University of Tennessee, Knoxville, Tennessee 37996, USA}
\affiliation{Min H. Kao Department of Electrical Engineering and Computer Science, University of Tennessee, Knoxville, Tennessee 37996, USA}

\author{Shi-Zeng Lin}%
\affiliation{Theoretical Division, T-4, Los Alamos National Laboratory, Los Alamos, New Mexico 87545, USA}
\affiliation{Center for Integrated Nanotechnologies (CINT), Los Alamos National Laboratory, Los Alamos, New Mexico 87545, USA}
\date{\today}

\begin{abstract}
The observation of superconductivity emerging from a spin- and valley-polarized
normal state in rhombohedral graphene raises a basic puzzle: intravalley Cooper
pairs involve states that are not
generically degenerate because of trigonal warping, and therefore lack the
particle--particle nesting underlying the conventional Cooper instability.
While recent model studies suggest that nontrivial band geometry can strongly
promote chiral pairing, how this geometric enhancement competes with realistic
band-structure-induced pair breaking remains unclear. Here we study
Kohn--Luttinger superconductivity in electron-doped rhombohedral tetralayer
graphene using a DFT-fitted multiorbital tight-binding model, retaining the full
Bloch-wave-function form factors in both RPA screening and the projected Cooper
interaction. We separate the superconducting instability into three
contributions: the symmetry-resolved form-factor-dressed interaction, the
Cooper-pair phase space determined by the symmetrized pair dispersion, and the
energy mismatch generated by trigonal warping. We find that the Bloch form
factors strongly favor intravalley pairing, with a leading predominantly
$p_x-ip_y$ instability and a substantially larger maximum $T_c$ than the
intervalley reference state. At the same time, trigonal-warping-induced
mismatch cuts off the Cooper logarithm and confines intravalley
superconductivity to a much narrower region of the phase diagram. The dominant
instability occurs near pair-Lifshitz transitions and primarily in reconstructed
annular or three-lobed Fermi-surface regimes. Our results show that
superconductivity in rhombohedral graphene is controlled by the combined
effects of quantum-geometry-dressed interactions, pair fermiology, and
trigonal-warping-induced pair breaking.
\end{abstract}

\maketitle

\section{Introduction}
Rhombohedral multilayer graphene has emerged as a highly tunable platform for exploring correlated topological quantum phases, combining narrow low-energy bands with strong Berry curvature and nontrivial quantum geometry~\cite{Kumar2013FlatBerry}. The interplay between band topology and strong interactions was spectacularly demonstrated in hBN-aligned rhombohedral pentalayer graphene, where fractional quantum anomalous Hall states---realizations of fractional Chern insulators---were observed in a moir\'e superlattice~\cite{Lu2024FQAH}. Remarkably, unconventional superconductivity was subsequently discovered in rhombohedral tetralayer and pentalayer graphene even in the absence of a moir\'e superlattice~\cite{Han2025ChiralSC}. These superconducting states emerge from a spin- and valley-polarized quarter-metal phase and exhibit spontaneous time-reversal-symmetry breaking and other signatures consistent with chiral superconductivity. These discoveries establish rhombohedral graphene as a unique setting in which strong correlations, quantum geometry, band topology, and unconventional superconductivity can be explored within the same electrically tunable material platform.

Following the initial observation of signatures of chiral superconductivity in rhombohedral tetralayer and pentalayer graphene~\cite{Han2025ChiralSC}, a rapidly growing body of experiments has revealed an unexpectedly rich superconducting phenomenology. 
Transport measurements have identified multiple superconducting states that are enhanced or even induced by magnetic fields, remain robust far beyond the conventional Pauli limit, and evolve strongly with carrier density, displacement field, layer number, and proximity-induced spin--orbit coupling~\cite{Seo2026FieldBoosted}. 
More recently, superconductivity with similar signatures of orbital time-reversal-symmetry breaking has been extended to rhombohedral hexalayer graphene, where magnetic field, density, and displacement field can switch between superconducting states associated with different spin--valley flavors~\cite{Hua2026Quartet}. 
Complementary nanoscale magnetic imaging has provided direct information on the underlying orbital magnetism: SQUID-on-tip measurements in pentalayer graphene have spatially resolved time-reversal-breaking superconducting domains and demonstrated electrical switching between states of opposite chirality~\cite{Dutta2026Reconfigurable}, while measurements over a broad range of layer numbers have imaged the orbital magnetization of the quarter-metal parent state and constrained the magnetic response of the superconducting phase~\cite{Sheekey2026OrbitalMagnetism}.  Such orbital magnetization in a quarter-metal metal has been
investigated theoretically and proposed to stabilize a spontaneous vortex
lattice in the superconducting state even in the absence of an applied
magnetic field~\cite{Jahin2025SpontaneousVortex}. 
At the same time, quantum-oscillation measurements have substantially refined---and complicated---our understanding of the normal state. 
In rhombohedral tetralayer graphene, measurements extending into the superconducting regime reveal a multitone oscillation spectrum incompatible with a simple, singly connected quarter-metal Fermi surface~\cite{Kalantre2026Fermiology}, whereas recent measurements in pentalayer graphene resolve an electrically tunable Lifshitz transition between simply connected and annular quarter-metal Fermi surfaces, with chiral superconducting states appearing on both sides of the transition~\cite{Zheng2026Lifshitz}. 
Taken together, these experiments place increasingly stringent constraints on the parent electronic state and on the broken symmetries of the superconducting phase, while at the same time raising fundamental questions concerning the microscopic origin of superconductivity ~\cite{Ghazaryan2023Multilayer,Chou2025Intravalley,Yang2025FFLO,ParraMartinez2025Band,Geier2026Chiral,Qin2026Chiral,Murshed2026RepulsivePDW,Chou2026Retardation,Shavit2025QuantumGeometric,Jahin2026Enhanced,Shavit2026Nematic,MayMann2026ValleyHelical,Murshed2025Nodal,Gil2026Charge,Yoon2026QuarterMetal,Christos2026FiniteMomentum,Wang2026Chiral,MayMann2026PairingTopology,Nguyen2026Kekule}. Despite this extensive theoretical activity, a unified microscopic understanding of which ingredients control the superconducting instability and determine its phase boundaries has yet to emerge.

One particularly important question is why intravalley superconductivity is favored in the first place. 
Unlike conventional intervalley pairing between time-reversed states, states at $K+\mathbf{k}$ and $K-\mathbf{k}$ on a single-valley Fermi surface are not generically degenerate because of trigonal warping and other band-structure asymmetries, and therefore do not possess perfect particle--particle nesting. 
Although the experimentally observed normal state is spin and valley polarized, this by itself does not guarantee a superconducting instability: valley polarization merely restricts the available low-energy degrees of freedom, while competing metallic, density-wave, or crystalline states may instead be stabilized. 
A possible resolution is provided by the nontrivial topology and quantum geometry of the active band. 
Model studies have shown that Bloch-wave-function form factors can strongly enhance Kohn--Luttinger pairing, with the Berry flux enclosed by the Fermi surface coupling to the angular momentum of Cooper pairs and producing a pronounced enhancement of chiral superconductivity~\cite{Jahin2026Enhanced,Shavit2025QuantumGeometric,MayMann2026PairingTopology}. 
This mechanism has been realized microscopically in the spin-polarized Chern band of twisted $\mathrm{MoTe}_2$, where realistic form factors favor intravalley chiral $p+ip$ pairing over the competing intervalley channel~\cite{Xu2025ChiralMoTe2}. 
Closely related physics has recently been identified in rhombohedral multilayer graphene itself: for suitable displacement fields, the low-energy conduction band develops a flat-bottom ``Berry trashcan'' structure whose projected-density form factors approximately obey the Girvin--MacDonald--Platzman algebra characteristic of a Landau level~\cite{bernevig2025berrytrashcanmodelinteracting,Jiang2026IdealQuantumGeometry}. 
Indeed, within this Landau-level-like model, short-range attraction admits exact $p_x\pm ip_y$ superconducting ground states whose chirality is tied to the Berry curvature~\cite{Li2025BerryTrashcanSC}. Quantum geometry can also directly influence the microscopic structure
of the superconducting state: the quantum metric and Berry curvature
have recently been shown to contribute to the Cooper-pair size in
rhombohedral graphene, with the Berry-curvature contribution becoming
particularly important when time-reversal symmetry is broken
~\cite{Chen2026QuantumGeometricQuadrupole}.
These results strongly suggest that band topology and quantum geometry can play an essential role in stabilizing intravalley chiral superconductivity. 
However, a material-specific understanding that treats realistic band dispersion, Bloch-wave-function form factors, Lifshitz transitions and the associated density of states, as well as trigonal-warping-induced pair breaking on equal footing remains lacking.

    \begin{figure}
        \centering
        \includegraphics[width=0.9\linewidth]{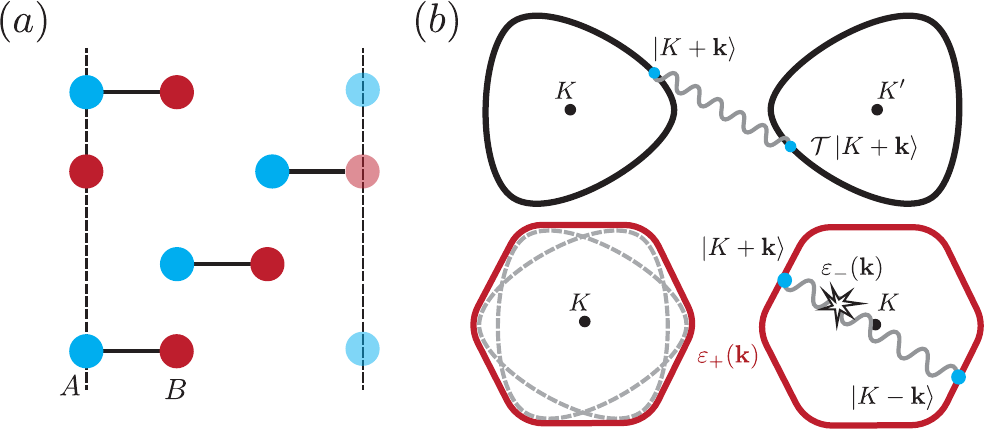}
        \caption{(a) Stacking structure of rhombohedral multilayer graphene. (b) Intervalley pairing between two time reversed states at the $K$ and $K'$ valley. In the case of intravalley pairing, the Fermi surface of $\varepsilon(K+\mathbf{k})$ and its inverse $\varepsilon(K-\mathbf{k})$ are averaged to give the pair $\varepsilon_+(\mathbf{k})$ which serves as the Fermi surface for the trigonally warped case. The mismatch between them, $\varepsilon_{-}(\mathbf{k})$ serves as a momentum dependent pair breaking field. }
        \label{fig:fig1}
    \end{figure}

In this work, we investigate the competition between intravalley and intervalley superconductivity in rhombohedral multilayer graphene. The two pairing channels face qualitatively different conditions. For intervalley pairing, electrons at $ \mathbf K+\mathbf{k}$ and $\mathbf K'-\mathbf{k}$ are related by time-reversal symmetry and are therefore naturally particle--particle nested. Intravalley pairing, by contrast, involves states at $\mathbf K+\mathbf{k}$ and $\mathbf K-\mathbf{k}$, which are not generically degenerate because of trigonal warping. This produces a momentum-dependent energy mismatch that cuts off the Cooper instability and acts as an effective pair-breaking field, depicted in Figure \ref{fig:fig1}. At the same time, intravalley pairing can benefit from the nontrivial Bloch-wave-function form factors associated with the quantum geometry of the active band, which modify both the screening of the Coulomb interaction and its projection into the Cooper channel. Moreover, because the relevant intravalley Cooper dispersion is the symmetrized combination of the two paired states, its Fermi surface and density of states can differ substantially from those of the underlying single-particle band. The relative stability of intra- and intervalley superconductivity is therefore governed by the interplay among the form-factor-dressed pairing interaction, the available Cooper-pair phase space, and trigonal-warping-induced pair breaking.

To quantify these effects in a material-specific setting, we construct realistic tight-binding models for rhombohedral graphene with different layer numbers using parameters fitted to density-functional-theory calculations, retaining the full multiorbital Bloch wave functions in both the RPA screening and the Cooper-pairing vertex. We phenomenologically polarize the system into the spin- and valley-polarized quarter-metal regime and solve the linearized gap equation over carrier density and displacement field. This framework allows us to separately characterize the symmetry-resolved effective pairing interaction, the single-particle and pair densities of states and their evolution across Lifshitz transitions, and the energy mismatch generated by trigonal warping. By combining these ingredients, we determine how quantum geometry, Fermi-surface topology, and band asymmetry cooperate or compete to shape the superconducting phase diagram and the competition between intravalley and intervalley pairing.

The remainder of this paper is organized as follows. In Sec.~II,
we classify the intra- and intervalley pairing states, and in Sec.~III
we introduce the realistic tight-binding model of rhombohedral graphene.
Sections~IV and V discuss the quantum geometry of the low-energy bands
and the Bloch-form-factor-dressed screened interaction.
In Sec.~VI, we formulate the pair Fermi surface and
trigonal-warping-induced pair breaking. Sections~VII--IX introduce
quantitative measures of the pair density of states, Lifshitz transitions,
pair-breaking strength, and symmetry-resolved pairing interactions.
Section~X describes our numerical methods. In Sec.~XI, we present
the superconducting phase diagrams and analyze the microscopic
mechanisms governing the pairing instabilities. Finally, Sec.~XII
discusses the implications for experiments and summarizes our conclusions.

\section{Rhombohedral Graphene and Symmetry of the Pairing States}\label{sec:formfactor_interaction}

Rhombohedral multilayer graphene consists of $N$ graphene layers stacked in the ABC sequence. In the chiral limit, the low-energy electronic states are predominantly localized on the two outermost non-dimer sublattices, and their dispersion near the two valleys $K$ and $K'$ scales approximately as $k^N$. As the number of layers increases, this produces increasingly flat conduction and valence bands near charge neutrality. A perpendicular displacement field breaks inversion symmetry, separates the low-energy conduction and valence bands, and generates strong valley-contrasting Berry curvature and nontrivial Bloch-wave-function geometry. Remote interlayer hoppings, most importantly those responsible for trigonal warping, reduce the approximate continuous rotational symmetry of the low-energy theory to the lattice $C_3$ symmetry and strongly reconstruct the Fermi surface. Depending on carrier density and displacement field, the low-energy band can consequently exhibit single-pocket, three-pocket, and annular Fermi surfaces separated by Lifshitz transitions.

In the absence of a displacement field, rhombohedral graphene has point group $D_{3d}$. A perpendicular displacement field breaks inversion symmetry and reduces the point group to $C_{3v}$. The superconducting order parameter, however, is classified according to the little group of the corresponding average Cooper-pair momentum. This distinction is important because intra- and intervalley pairing are centered at different momenta and therefore transform under different symmetry groups.

We label the two valleys by $\tau=\pm$, corresponding to $K$ and $K'$, and measure $\mathbf{k}$ relative to the valley center. Intravalley pairing in valley $\tau$ involves $(K_{\tau}+\mathbf{k},\ K_{\tau}-\mathbf{k})$,
and we define the corresponding pair amplitude as $\Delta_{\tau\tau}(\mathbf{k})=\left\langle c_{\tau,\mathbf{k}}c_{\tau,-\mathbf{k}}\right\rangle$.
Because a single valley is not invariant under the mirror operations that exchange $K$ and $K'$, the relevant little group for a valley-polarized intravalley state is $C_3$. The three one-dimensional irreducible representations of $C_3$ can be labeled by $r=0,\,+1,\,-1$, according to their eigenvalue under a threefold rotation, $\Delta_r(R_3\mathbf{k})=e^{i2\pi r/3}\Delta_r(\mathbf{k})$.
In the approximate continuous-rotation limit, a gap function with angular dependence $\Delta_l(\theta)\propto e^{il\theta}$
belongs to the $C_3$ sector $r=l\pmod 3$. Consequently, angular-momentum channels that differ by multiples of three can mix once the full lattice symmetry is taken into account.

For the spin- and valley-polarized quarter metal, the active electrons are effectively single-component fermions. Fermi statistics therefore require $\Delta_{\tau\tau}(-\mathbf{k})=-\Delta_{\tau\tau}(\mathbf{k})$, 
which restricts the superconducting order parameter to odd angular momentum in the rotationally symmetric limit. The three $C_3$ sectors then contain, for example,
\begin{align}
    r=+1 &: \qquad
    l=\ldots,-5,1,7,\ldots,
    \nonumber\\
    r=-1 &: \qquad
    l=\ldots,-7,-1,5,\ldots,
    \nonumber\\
    r=0 &: \qquad
    l=\ldots,-9,-3,3,9,\ldots .
    \label{eq:C3_angular_channels}
\end{align}
The lowest angular harmonics therefore correspond to the two chiral $p$-wave states,
\begin{equation}
    p_x+ip_y
    \sim e^{i\theta},
    \qquad
    p_x-ip_y
    \sim e^{-i\theta},
\end{equation}
and the $f$-wave channel, $f\sim e^{\pm 3i\theta}$. Because time-reversal symmetry is already broken in a valley-polarized normal state, the $r=+1$ and $r=-1$ representations are not required to be degenerate.

Intervalley pairing has a different symmetry structure. We consider a single spinless band. In this case the two electrons are taken from opposite valleys, $(\mathbf K+\mathbf{k},\ \mathbf K'-\mathbf{k})$, and the pair amplitude is $\Delta_{+-}(\mathbf{k})=\left\langle c_{+,\mathbf{k}} c_{-,-\mathbf{k}}\right\rangle$. Since $\mathbf K+\mathbf K'$ is equivalent to zero crystal momentum, intervalley pairing is centered at the $\Gamma$ point. Its orbital structure is therefore classified by the full $C_{3v}$ point group of the displacement-field-broken normal state. The relevant irreducible representations are $A_1,\qquad A_2,\qquad E$.
In the approximate continuous-rotation limit in the coordinate system of $\mathbf{k}$, i.e. $\theta$ measured from the valley center, representative basis functions are
\begin{align}
    A_1 &: \qquad \cos 3\theta, \ \sin 6\theta, \ \cos 9\theta \ ,
    \ldots 
    \nonumber\\
    A_2 &: \qquad
    1, \ \sin 3\theta, \ \cos 6\theta, \ldots 
    \nonumber\\
    E &: \qquad
    \left\{
    \cos l\theta,\,
    \sin l\theta
    \right\},
    \qquad
    l\neq 0\pmod 3 .
    \label{eq:C3v_basis}
\end{align}

%
%
%
For even $l$ we have valley-singlet and for odd $l$ we have valley-triplet. A constant order parameter in one valley belongs to $A_2$ while the two components $(p_x,p_y)$ transform as the two-dimensional $E$ representation. Higher angular harmonics with the same lattice quantum numbers can mix in a realistic band structure.

The two components of the $E$ representation are exactly degenerate at the level of the linearized gap equation as long as $C_{3v}$ symmetry is preserved. The quadratic theory therefore determines only that the instability belongs to the $E$ representation; it does not distinguish between a real nematic combination and a chiral combination such as $p_x\pm ip_y$. This distinction is determined by quartic terms in the Ginzburg--Landau free energy below $T_c$.

For intervalley pairing, Fermi antisymmetry involves both momentum and valley indices, $\Delta_{+-}(\mathbf{k})=-\Delta_{-+}(-\mathbf{k})$.
Consequently, the parity of the orbital wave function is tied to the symmetry of the pair under valley exchange (and, when spin is retained explicitly, to its spin-singlet or spin-triplet structure). Both even- and odd-angular-momentum orbital representations can therefore arise in the intervalley problem, in contrast to the effectively spinless intravalley quarter-metal state where only odd-parity pairing is allowed.

The symmetry distinction between the two pairing channels will be important below. For intravalley pairing, we project the effective interaction onto the three $C_3$ sectors $r=0,\pm1$, whereas for intervalley pairing we resolve it into the $C_{3v}$ representations $A_1$, $A_2$, and $E$. This symmetry-resolved formulation allows us to compare the pairing interactions themselves independently of the density of states and of trigonal-warping-induced pair breaking.

\section{Model}

We describe rhombohedral $4$-layer graphene using a Slonczewski--Weiss--McClure (SWM) parameterized tight-binding Hamiltonian~\cite{PhysRev.108.612,PhysRev.109.272,PhysRevB.82.035409}, with parameters obtained by fitting the low-energy $p_z$ bands to density-functional-theory calculations performed using the \textsc{Quantum ESPRESSO} package~\cite{PhysRevLett.77.3865,GARRITY2014446,Giannozzi_2009,Giannozzi_2017}. 

To model the spin- and valley-polarized quarter-metal regime, we work with an effectively spinless model and phenomenologically introduce a valley-polarizing next-nearest-neighbor hopping,
\begin{equation}
H_{\mathrm{vp}}
=
\gamma
\sum_{\mathbf{k},\boldsymbol{\delta}_i}
\left[
\sin(\mathbf{k}\cdot\boldsymbol{\delta}_i)
I_N\sigma_0
\right]
c_{\mathbf{k}\alpha}^{\dagger}
c_{\mathbf{k}\alpha},
\label{eq:valley_polarization}
\end{equation}
where the vectors $\boldsymbol{\delta}_i$ are related by $C_3$ rotations, $I_N$ acts in layer space, $\sigma_0$ is the identity in sublattice space, and $\gamma$ controls the strength of the valley splitting. This term shifts one valley away from the low-energy sector and provides a simple phenomenological description of the quarter-metal state relevant to intravalley pairing. The perpendicular displacement field is incorporated as a layer-dependent onsite potential. In our further discussions we set $\gamma$ independently at each value of carrier density and displacement field so that  the Fermi level sits at midgap at the $K'$ point. 

Electron-electron interactions are described by a dual-gate-screened Coulomb potential. Denoting the distance from the graphene multilayer to the gates by $d_{\rm g}$, the interaction between electrons at vertical coordinates $z$ and $z'$ is
\begin{equation}
V(q,z,z')
=
\frac{2\pi e^2}{\epsilon q}
\frac{
\cosh q(2d_{\rm g}-|z-z'|)
-
\cosh q(z+z')
}{
\sinh(2qd_{\rm g})
}.
\label{eq:gate_screened_coulomb}
\end{equation}
Projecting Eq.~\eqref{eq:gate_screened_coulomb} onto the localized $p_z$ orbitals gives the orbital-resolved interaction  $V_{\alpha\beta}(\mathbf q)$. Unless otherwise stated, we use a relative dielectric constant $\varepsilon_r=6$ and a gate distance $d_{\rm g}=200~\AA$. 

We study both intravalley and intervalley superconducting instabilities. Intravalley pairing occurs between states $K+\mathbf{k}$ and $K-\mathbf{k}$, and therefore has center-of-mass momentum at the valley center $K$ (or equivalently $K'$). For this calculation we retain the phenomenological valley-polarizing term in Eq.~\eqref{eq:valley_polarization}, so that only one valley remains active near the Fermi energy. For comparison, we also calculate the intervalley instability between the time-reversed states $\mathbf{k}$ and $-\mathbf{k}$, whose total crystal momentum lies at $\Gamma$. In the intervalley calculation we set $\gamma=0$ and restore the valley-degenerate band structure. Thus the intervalley calculation should be viewed as a reference instability of the corresponding valley-unpolarized system rather than as a competing instability of precisely the same valley-polarized Hamiltonian. In the intravalley case we neglect Umklapp effects and consider only the lowest wavevector $\mathbf{G}$. In the intervalley case we consider two Umklapp vectors so that the interaction matrix elements contain all $C_3$ related momentum transfers. 

The resulting superconducting problem is treated within a Fermi-surface formulation of the linearized gap equation. This formulation makes it possible to separate three effects that vary strongly across the phase diagram: the form-factor-dressed pairing interaction, the available Cooper-pair phase space, and the energy mismatch between the two members of an intravalley Cooper pair.

\section{Quantum Geometry of Rhombohedral Graphene}
\label{sec:quantum_geometry}
The discussion above focused primarily on the electronic dispersion and the associated
Fermi-surface structure. For superconductivity, however, the single-particle Bloch wave
functions are equally important, since they determine how the interaction is projected into
the low-energy band and therefore directly enter the Cooper-pairing vertex. In contrast to
many semiclassical responses that are controlled by the local properties of the Bloch states,
superconducting scattering generally connects states separated by finite, and potentially
large, momentum transfers. Consequently, local geometric quantities such as the Berry
curvature and quantum metric do not by themselves provide a complete characterization of
the form factors relevant to pairing. Nevertheless, they provide a natural starting point for
understanding how the geometry and topology of the Bloch wave functions influence the
superconducting instability and for identifying regimes in which the band approaches
Landau-level-like quantum geometry. 

The geometry of an isolated Bloch band is characterized by the quantum geometric
tensor~\cite{ProvostVallee1980,Roy2014BandGeometry},
\begin{equation}
    \mathcal{Q}_{ij}(\mathbf{k})
    =
    \left\langle
    \partial_{k_i}u_{\mathbf{k}}
    \left|
    1-|u_{\mathbf{k}}\rangle\langle u_{\mathbf{k}}|
    \right|
    \partial_{k_j}u_{\mathbf{k}}
    \right\rangle ,
    \label{eq:QGT}
\end{equation}
whose real and imaginary parts define the quantum metric and Berry curvature,
respectively,
\begin{equation}
    g_{ij}(\mathbf{k})
    =
    \operatorname{Re}\mathcal{Q}_{ij}(\mathbf{k}),
    \qquad
    \Omega_{ij}(\mathbf{k})
    =
    -2\,\operatorname{Im}\mathcal{Q}_{ij}(\mathbf{k}).
    \label{eq:metric_curvature}
\end{equation}

The quantum metric and Berry curvature obey the trace
inequality~\cite{Roy2014BandGeometry}
\begin{equation}
    \operatorname{Tr}g(\mathbf{k})
    \equiv
    g_{xx}(\mathbf{k})+g_{yy}(\mathbf{k})
    \geq
    |\Omega(\mathbf{k})|,
    \label{eq:trace_inequality}
\end{equation}
where $\Omega\equiv\Omega_{xy}$.

For the chiral model of rhombohedral $N$-layer graphene, it is convenient to work in the layer--sublattice basis
$|\alpha,l\rangle$, where $\alpha=A,B$ labels the sublattice and
$l=1,\ldots,N$ labels the layer. In the chiral limit, retaining only the
intralayer Dirac hopping $v_F$ and the dominant nearest-layer hopping
$t_1$, the low-energy non-dimer sector is spanned by two orthogonal
chiral states~\cite{bernevig2025berrytrashcanmodelinteracting},
\begin{align}
    [\Psi_A(\mathbf{k})]_{\alpha l}
    &=
    \frac{1}{\mathcal N(\mathbf{k})}
    \left(1,x,\ldots,x^{N-1}\right)_l
    \delta_{\alpha A},
    \\
    [\Psi_B(\mathbf{k})]_{\alpha l}
    &=
    \frac{1}{\mathcal N(\mathbf{k})}
    \left(\bar{x}^{N-1},\ldots,\bar{x},1\right)_l
    \delta_{\alpha B},
\end{align}
where $x=-\frac{v_F}{t_1}k_-$, $k_\pm=k_x\pm i k_y$, up to convention-dependent factors associated with the definition of the
continuum momentum. The states $\Psi_A$ and $\Psi_B$ are holomorphic and
antiholomorphic, respectively, and for $|x|<1$ are exponentially
localized toward opposite outer surfaces of the rhombohedral stack.
They become exact low-energy states in the $\mathbf{k}\rightarrow0$
limit. A displacement field selects predominantly one of these two
chiral states for the low-energy conduction band. In this basis, the band form factor then reduces to
\begin{equation}
    \Lambda_{\mathbf{k},\mathbf{k}+\mathbf{q}}
    \simeq
    \exp\left[
    -\frac{\beta q^2}{2}
    -i\beta\,\mathbf q\times\mathbf k
    \right],
    \qquad
    \beta=\frac{v_F^2}{t_1^2}.
    \label{eq:chiral_graphene_formfactor}
\end{equation}
This has the same structure as the lowest-Landau-level form factor, providing the microscopic origin of the Landau-level-like quantum geometry of the chiral rhombohedral graphene band. In this limit, the Berry curvature and quantum metric are both momentum independent, $|\Omega|=2\beta$, $g_{ij}=\beta\,\delta_{ij}$. They therefore saturate the ideal trace condition, $\operatorname{Tr}g=|\Omega|$,
showing that the chiral band realizes the same ideal quantum geometry as the lowest Landau level in this approximation.

Below we compute
$\Omega(\mathbf{k})$, $\operatorname{Tr}g(\mathbf{k})$, and the deviation from the ideal trace condition,
\begin{equation}
    \delta_{\rm tr}(\mathbf{k})
    =
    (\operatorname{Tr}g(\mathbf{k})
    -
    |\Omega(\mathbf{k})|)/\operatorname{Tr}g(\mathbf{k}),
\end{equation}
which vanishes for ideal Landau-level-like quantum geometry. We correlate these
quantities with the intra- and intervalley superconducting instabilities in
order to determine how proximity to Landau-level-like quantum geometry affects
the pairing interaction.

\section{Screened Interaction and Bloch Form Factors}

We consider a Kohn--Luttinger mechanism in which the repulsive gate-screened Coulomb interaction is further renormalized by particle-hole fluctuations. Within the random phase approximation (RPA), we first evaluate the static susceptibility in the orbital basis,
\begin{align}
\chi^{0}_{\alpha\beta}(\mathbf q)
=
\int\frac{d^2\mathbf k}{(2\pi)^2}
\sum_{mn}
&
\frac{
n_F(\varepsilon_{m\mathbf k})
-
n_F(\varepsilon_{n,\mathbf{k+q}})
}{
\varepsilon_{m\mathbf k}
-
\varepsilon_{n,\mathbf{k+q}}
}
\nonumber\\
&\times
u_{\alpha n}(\mathbf{k+q})
u^{*}_{\alpha m}(\mathbf k)
u^{*}_{\beta n}(\mathbf{k+q})
u_{\beta m}(\mathbf k).
\label{eq:susceptibility}
\end{align}
The screened interaction is then obtained from the matrix RPA resummation,
\begin{equation}
W(\mathbf q)
=
\left[
1-V(\mathbf q)\chi^{0}(\mathbf q)
\right]^{-1}
V(\mathbf q).
\label{eq:RPA_interaction}
\end{equation}

For superconductivity with pairing center $\mathbf K$, the relevant Cooper process scatters $(\mathbf K+\mathbf k,\mathbf K-\mathbf k)
\longrightarrow
(\mathbf K+\mathbf k',\mathbf K-\mathbf k')$.
Projection of the screened interaction onto band $m$ gives
\begin{align}
\widetilde V(\mathbf k,\mathbf k')
=
\sum_{\alpha\beta}
&
u^{*}_{m\alpha}(\mathbf K+\mathbf k')
u_{\alpha m}(\mathbf K+\mathbf k)
\nonumber\\
&\times
u^{*}_{m\beta}(\mathbf K-\mathbf k')
u_{\beta m}(\mathbf K-\mathbf k)
W_{\alpha\beta}(\mathbf k'-\mathbf k).
\label{eq:projected_pairing_interaction}
\end{align}
Equation~\eqref{eq:projected_pairing_interaction} makes clear that the Bloch wave functions affect the effective Cooper interaction in two distinct ways. First, their orbital structure enters the particle-hole susceptibility and therefore modifies the RPA screening in Eq.~\eqref{eq:RPA_interaction}. Second, the Bloch overlaps enter directly in the projection of the screened interaction into the Cooper channel. Both contributions are retained in our calculations.

The distinction between intra- and intervalley pairing is particularly transparent at the level of the Cooper form factors. For intravalley pairing, the two overlap factors take the schematic form
\begin{equation}
F_{\rm intra}(\mathbf k,\mathbf k')
=
\Lambda_{K}(\mathbf k,\mathbf k')
\Lambda_{K}(-\mathbf k,-\mathbf k'),
\label{eq:intra_formfactor}
\end{equation}
where $\Lambda_K(\mathbf k,\mathbf k')=\langle u_{K+\mathbf k'}|u_{K+\mathbf k}\rangle$.

Because both electrons belong to the same valley, the complex phase structure of the Bloch wave functions is retained in $F_{\rm intra}$. This allows Berry curvature and other aspects of quantum geometry to modify the angular structure of the Cooper interaction.

For intervalley pairing, by contrast, time-reversal symmetry relates the two valleys, $\Lambda_{K'}(-\mathbf k,-\mathbf k')=\Lambda_K^{*}(\mathbf k,\mathbf k')$, and therefore
\begin{equation}
F_{\rm inter}(\mathbf k,\mathbf k')
=
\Lambda_K(\mathbf k,\mathbf k')
\Lambda_{K'}(-\mathbf k,-\mathbf k')
=
\left|
\Lambda_K(\mathbf k,\mathbf k')
\right|^2.
\label{eq:inter_formfactor}
\end{equation}
The intervalley Cooper form factor is consequently real and non-negative. The complex phase winding of the Bloch wave functions that can distinguish different chiral intravalley channels is absent from the intervalley vertex, although the magnitude of the form factors continues to affect both screening and the strength of the projected interaction.

\section{Trigonal Warping and Effective Pair Breaking}

A central difference between intra- and intervalley pairing arises from particle-particle nesting. For intervalley pairing, the states $\mathbf k$ and $-\mathbf k$ are related by time-reversal symmetry and are therefore degenerate in the valley-unpolarized normal state. For intravalley pairing, however, no symmetry generally requires the states $\mathbf K+\mathbf k$ and $\mathbf K-\mathbf k$ to have the same energy. Trigonal warping therefore produces an intrinsic energy mismatch between the two members of an intravalley Cooper pair.

To make this explicit, we introduce the symmetric and antisymmetric combinations
\begin{equation}
\varepsilon_{\pm}(\mathbf k)
=
\frac{1}{2}
\left[
\varepsilon(\mathbf K+\mathbf k)
\pm
\varepsilon(\mathbf K-\mathbf k)
\right].
\label{eq:eps_pm}
\end{equation}
The particle-particle propagator entering the linearized gap equation can then be written as
\begin{align}
k_BT
\sum_n
&G(i\omega_n,\mathbf K+\mathbf k)
G(-i\omega_n,\mathbf K-\mathbf k)
\nonumber\\
&=
k_BT
\sum_n
\frac{1}{
\left[
\omega_n+i\varepsilon_-(\mathbf k)
\right]^2
+
\varepsilon_+^2(\mathbf k)
}.
\label{eq:particle_particle_kernel}
\end{align}
This decomposition has a direct physical interpretation. The symmetric dispersion $\varepsilon_+(\mathbf k)$ determines the kinematics and phase space of Cooper pairing, whereas $\varepsilon_-(\mathbf k)$ measures the energy mismatch within a pair and therefore acts as a momentum-dependent pair-breaking field.

The natural Fermi surface for the intravalley Cooper problem is consequently not the ordinary single-particle Fermi surface, but the \emph{pair Fermi surface} defined by
\begin{equation}
\varepsilon_+(\mathbf k)=0.
\label{eq:pair_fs}
\end{equation}
Parameterizing this contour by $s$ and using $\xi=\varepsilon_+$ as the coordinate normal to it, the integration measure is $d^2k=\frac{ds\,d\xi}{|\nabla_{\mathbf k}\varepsilon_+|}$ and the linearized gap equation becomes, in the constant interaction, gap, and density of states approximation, 
\begin{equation}
\Delta(s)
=
-
\oint_{\rm PFS}
\frac{ds'}{(2\pi)^2
|\nabla_{\mathbf k}\varepsilon_+(s')|}
\,
\widetilde V(s,s')
K(s',T)
\Delta(s'),
\label{eq:FS_gap}
\end{equation}
where
\begin{equation}
K(s,T)
=
k_BT
\sum_n
\int_{-W_l}^{W_u}
d\xi\,
\frac{1}{
[\omega_n+i\varepsilon_-(s)]^2+\xi^2
}.
\label{eq:pair_kernel}
\end{equation}
Here $W_l$ and $W_u$ denote the available pairing bandwidths below and above the pair Fermi level, respectively. For a symmetric and sufficiently large cutoff, $W_l=W_u=W\gg k_BT,\ |\varepsilon_-|$,
Eq.~\eqref{eq:pair_kernel} reduces to
\begin{align}
K(s,T)
=
&
\ln\left(
\frac{2e^{\gamma}W}{\pi k_BT}
\right)
\nonumber\\
&-
\left[
\operatorname{Re}
\psi\left(
\frac{1}{2}
+
\frac{i\varepsilon_-(s)}
{2\pi k_BT}
\right)
-
\psi\left(\frac{1}{2}\right)
\right],
\label{eq:Wbig}
\end{align}
where $\psi$ is the digamma function. The first term is the usual BCS Cooper logarithm. The second term is positive and suppresses the pairing susceptibility whenever $\varepsilon_-(s)\neq0$. The structure is analogous to standard pair-breaking problems~\cite{osti_4097498}, except that here the pair-breaking scale varies around the pair Fermi surface.

An important consequence of Eq.~\eqref{eq:Wbig} is that imperfect
particle--particle nesting produces a finite critical interaction strength.
For $T\rightarrow0$,
\begin{equation}
    \operatorname{Re}\psi\left(
    \frac{1}{2}
    +\frac{i\varepsilon_-}{2\pi k_BT}
    \right)
    \simeq
    \ln\left|
    \frac{\varepsilon_-}{2\pi k_BT}
    \right|,
\end{equation}
and hence $K(s,0)=\ln\frac{W}{|\varepsilon_-(s)|}$. Thus the energy mismatch cuts off the Cooper logarithm even at zero
temperature. For a momentum-independent attractive interaction
$\widetilde V=-V_0$, the superconducting instability exists only when $V_0>V_c$,
where
\begin{equation}
    V_c^{-1}
    =
    \oint_{\rm PFS}
    \frac{ds}{(2\pi)^2|\nabla_{\mathbf k}\varepsilon_+|}
    \ln\frac{W}{|\varepsilon_-(s)|}.
    \label{eq:critical_interaction}
\end{equation}
Equivalently, defining $\lambda=V_0N_{\rm pair}$,
\begin{equation}
    \lambda_c
    =
    \left[
    \left\langle
    \ln\frac{W}{|\varepsilon_-|}
    \right\rangle_{\rm PFS}
    \right]^{-1}.
    \label{eq:critical_lambda}
\end{equation}
For a momentum-independent mismatch
$|\varepsilon_-|=h$, this reduces to $\lambda_c=\frac{1}{\ln(W/h)}$.
Hence perfect particle--particle nesting, $h\rightarrow0$, recovers the
usual Cooper instability at arbitrarily weak attraction, whereas any finite
mismatch cuts off the logarithmic divergence and requires a finite pairing
strength.

Close to a band edge, however, the constant-density-of-states and large-symmetric-cutoff approximations become unreliable. In the opposite limit $W_l=W_u=W\ll k_BT$, one obtains
\begin{equation}
K(s,T)
\simeq
\frac{W}{\pi^2 k_BT}
\operatorname{Re}
\psi^{(1)}
\left(
\frac{1}{2}
+
\frac{i\varepsilon_-(s)}
{2\pi k_BT}
\right),
\label{eq:Wsmall}
\end{equation}
where $\psi^{(1)}$ is the first polygamma function. Since
$\psi^{(1)}(z)\sim z^{-1}$ at large $|z|$, the logarithmic Cooper enhancement is lost when the accessible pairing bandwidth becomes comparable to or smaller than the temperature. In the numerical calculations we therefore evaluate Eq.~\eqref{eq:pair_kernel} directly rather than relying on either asymptotic limit. The cutoff on the side closest to the band edge is determined from the corresponding edge of the symmetrized dispersion, while the opposite cutoff is fixed at $2.5~\mathrm{meV}$. 

This formulation reveals a second important consequence of intravalley pairing. The phase-space factor entering Eq.~\eqref{eq:FS_gap} is determined by the symmetrized dispersion,
\begin{equation}
N_{\rm pair}(0)
=
\oint_{\rm PFS}
\frac{ds}{
(2\pi)^2
|\nabla_{\mathbf k}\varepsilon_+|
},
\label{eq:pair_DOS_short}
\end{equation}
rather than by the ordinary single-particle density of states. Consequently, van Hove singularities and Lifshitz transitions of the pair dispersion need not coincide with those of the underlying electronic band. In particular, an ordinary Fermi surface can remain present even when Eq.~\eqref{eq:pair_fs} has no solution, in which case the weak-coupling intravalley Cooper instability is absent.

The intravalley superconducting instability is therefore controlled by three conceptually distinct ingredients. The Bloch form factors determine the symmetry-resolved effective pairing interaction; the symmetrized dispersion determines the available Cooper-pair phase space; and the antisymmetric dispersion produces trigonal-warping-induced pair breaking. In the following we quantify these effects separately before combining them in the full linearized gap equation. This decomposition allows us to identify which microscopic ingredient controls the location, symmetry, and strength of superconductivity across the phase diagram.

\section{Density of States and Lifshitz Transitions}

To quantify the phase space available for superconductivity, we distinguish the ordinary single-particle density of states from the pair density of states introduced above. At the Fermi level,
\begin{equation}
N_{\rm sp}
=
\sum_{\rm FS}
\oint_{\rm FS}
\frac{ds}
{(2\pi)^2|\nabla_{\mathbf k}\varepsilon|},
\label{eq:DOS_sp}
\end{equation}
where the sum runs over all Fermi-surface pockets. For intravalley pairing, the relevant phase-space measure is instead $N_{\rm pair}$ defined in Eq. \eqref{eq:pair_DOS_short}. The latter is precisely the density-of-states factor appearing in the Fermi-surface-restricted gap equation.

We evaluate both quantities as functions of carrier density and displacement field, using the corresponding chemical potential $\mu(n,D)$. We also track changes in the topology of both the ordinary Fermi surface and the pair Fermi surface. These define, respectively, single-particle and pair Lifshitz transitions, which need not occur at the same $(n,D)$. In particular, the pair Fermi surface may undergo a topology change, or disappear altogether, while the ordinary Fermi surface remains finite or continuous.

In the weak-coupling limit, neglecting pair breaking and assuming a slowly varying density of states, the dimensionless pairing strength $\lambda_\alpha
\sim
\mathcal A_\alpha N_{\rm pair}$, where $\mathcal A_\alpha$ denotes the attractive interaction strength defined below. Thus $N_{\rm pair}$ provides a direct measure of the Cooper-pair phase space. Close to a van Hove singularity or band edge, where the density of states varies rapidly within the pairing window, we retain the Fermi surface component of its momentum dependence in the numerical gap equation rather than relying on $\lambda_\alpha$.

\section{Quantifying Trigonal-Warping Pair Breaking}\label{sec:trigonal_warping}

The antisymmetric dispersion $\varepsilon_-(\mathbf k)$ introduced above provides a local measure of the energy mismatch between the two members of an intravalley Cooper pair. To characterize its overall strength by a single normal-state quantity, we define the pair-DOS-weighted root-mean-square mismatch
\begin{equation}
\Gamma_{\rm tw}
=
\left[
\frac{
\displaystyle
\oint_{\rm PFS}
\frac{ds}{|\nabla_{\mathbf k}\varepsilon_+|}
\,\varepsilon_-^2(\mathbf k)
}{
\displaystyle
\oint_{\rm PFS}
\frac{ds}{|\nabla_{\mathbf k}\varepsilon_+|}
}
\right]^{1/2}.
\label{eq:Gamma_tw}
\end{equation}
The weighting is the same as that appearing in the pair density of states, so that $\Gamma_{\rm tw}$ measures the characteristic mismatch experienced by states that contribute to the Cooper instability.

The connection to $T_c$ follows directly from the pair-breaking kernel in Eq.~\eqref{eq:Wbig}. For $|\varepsilon_-|\ll k_BT_c$, the digamma function gives
\begin{equation}
\operatorname{Re}
\psi\left(
\frac{1}{2}
+
\frac{i\varepsilon_-}{2\pi k_BT}
\right)
-
\psi\left(\frac{1}{2}\right)
\simeq
\frac{7\zeta(3)}{4\pi^2}
\frac{\varepsilon_-^2}{(k_BT)^2}.
\end{equation}
If the interaction and gap function vary slowly around the pair Fermi surface, averaging with the pair-DOS measure yields
\begin{equation}
\ln\frac{T_{c0}}{T_c}
\simeq
\frac{7\zeta(3)}{4\pi^2}
\left(
\frac{\Gamma_{\rm tw}}{k_BT_c}
\right)^2,
\label{eq:Tc_Gamma_tw}
\end{equation}
where $T_{c0}$ is the transition temperature obtained from the same pairing problem after setting $\varepsilon_-=0$. For weak suppression,
\begin{equation}
\frac{T_{c0}-T_c}{T_{c0}}
\simeq
0.213
\left(
\frac{\Gamma_{\rm tw}}{k_BT_{c0}}
\right)^2.
\label{eq:Tc_Gamma_tw_weak}
\end{equation}
Thus the dimensionless ratio $\frac{\Gamma_{\rm tw}}{k_BT_{c0}}$ provides a convenient measure of the strength of trigonal-warping-induced pair breaking. In the full calculation we retain the complete momentum dependence of $\varepsilon_-(\mathbf k)$; $\Gamma_{\rm tw}$ is used only as a compact diagnostic across the phase diagram.

\section{Quantifying the Form-Factor Contribution to Pairing}

To isolate the effect of Bloch-wave-function geometry from the density of states and pair breaking, we characterize the projected interaction independently of the particle-particle susceptibility. On the relevant pairing contour $\mathcal C$, we define
\begin{equation}
[\hat{\mathcal V}f](\mathbf k)
=
\oint_{\mathcal C}
\frac{ds'}{L_{\mathcal C}}\,
\widetilde V(\mathbf k,\mathbf k')
f(\mathbf k'),
\qquad
L_{\mathcal C}
=
\oint_{\mathcal C}ds .
\label{eq:interaction_only_kernel}
\end{equation}
The use of the normalized arclength measure, rather than the density-of-states weight
$ds/|\nabla\varepsilon|$, ensures that the resulting eigenvalues characterize the interaction itself. The corresponding discrete operator is $V(s, s') \sqrt{ds ds'}/L$ which we use in the following. 

For intravalley pairing, we project Eq.~\eqref{eq:interaction_only_kernel} onto the three $C_3$ sectors $r=0,\ \pm1$, while for intervalley pairing we project onto the $C_{3v}$ irreducible representations $\alpha=A_1,\ A_2,\ E$.
For each symmetry sector we define the interaction strength from its most attractive eigenvalue,
\begin{equation}
\mathcal A_\alpha
=
\max\left[
0,
-\min\operatorname{Spec}
\left(
\hat P_\alpha
\hat{\mathcal V}
\hat P_\alpha
\right)
\right].
\label{eq:A_alpha}
\end{equation}
Here $\hat P_\alpha$ denotes the appropriate $C_3$ or $C_{3v}$ symmetry projector.

\section{Numerical methods}

The main cost of the computation is the construction of the temperature dependent $\chi_{mn}(\mathbf{q})$ for every momentum transfer $\mathbf{q}$ connecting pair Fermi surface points. We solve the linearized gap equation together with the RPA interaction self consistently in temperature at fixed $\mu$ to a tolerance of 0.1 K. In the intervalley case the temperature is further estimated by the BCS formula. In order to speed up the computation, the susceptibility is interpolated between fixed constructions at 0, 2, 5, 7, and 10 K. At 0~K the susceptibility is integrated on an adaptive quad mesh refined around the $K$-point using the linear tetrahedron (triangle) method. At 2~K and above, a 7-point Gauss quadrature rule over a triangle is used \cite{cubature_rule}. The adaptive mesh is necessary in order to capture the fine details at small momentum transfers. The small size of the Fermi surface means that the susceptibility is not necessarily completely dominated by transitions near the Fermi surface, and hence a coarse mesh is used to capture contributions from the smooth interband background. To converge this integral we use between 7000 and 53000 $k$-points depending on the pocket size. This converges the $\chi$ integral to a relative tolerance of about $10^{-3}$ which is on the order of the interpolation error in temperature. This is roughly doubled in the intervalley case since we need to resolve both valleys. For the construction of $\chi$ for the intervalley case we consider the sum of the $K-K'$ momentum transfer and its two $C_3$ related Umklapp transfers.

As mentioned before, the asymptotic forms of the kernel derived above are unreliable whenever $\varepsilon_{-} \sim W$, and are hence summed numerically. To speed up the evaluation of the Matsubara sums, we replace the large-$n$ summations with their asymptotic forms. Here we consider the symmetric pairing window $[-W, W]$ for simplicity of notation. There are three scales in this problem: the dimensionless cutoff $M = W/2\pi k_B T$, the dimensionless pair breaking field $\tilde \varepsilon_- = \varepsilon_{-}/2\pi k_B T$, and the summation spacing $1$. We need to calculate
\begin{align*}
K(s, T) = \frac{1}{\pi}\sum_{n}\arctan\left(\frac{M}{n + 1/2 + i\tilde \varepsilon_{-}(s)} \right)\frac{1}{n + 1/2 + i\tilde \varepsilon_{-}(s)}.
\end{align*}

The two asymptotic limits for $M$ were discussed above. Whenever $M - |\tilde \varepsilon | > 5$ (Matsubara spacings) we directly use the large $M$ result. In the other case $M - |\tilde \varepsilon | < 5$ the first $N = \max(64, \textrm{ceil}(4M)+8) $ terms are directly summed and the remaining terms from $N+1$ to $\infty$ are evaluated by taking the expansion of $\arctan(z)=\sum_{m=0}^{\infty}(-1)^m z^{2m+1}/(2m+1)$ and keeping 9 terms. Each term can then be summed in terms of the Hurwitz zeta function $\zeta(s, a) = \sum_{n=0}^{\infty} (n+a)^{-s}$, which is evaluated using the mpmath library \cite{mpmath}. 

To discretize the Fermi surface efficiently, we uniformly sample with respect to a special metric. In the intervalley case, we sample uniformly with respect to the density of states measure $ds / \nabla \varepsilon$. In the intravalley case we include the asymptotic form of the pair breaking kernel, so that the spacing is decided by the measure $K ds / \nabla \varepsilon_{+}$ calculated at 1 K. This improves convergence with respect to the number of points used for the Fermi surface, especially near Lifshitz transitions. 

We performed density functional calculations in Quantum Espresso 7.3 \cite{Giannozzi_2009, Giannozzi_2017} using the PBE functional \cite{PhysRevLett.77.3865} with ultrasoft pseudopotentials \cite{GARRITY2014446} on a $48\times48$ mesh. The interlayer spacing was fixed at 3.348 \AA \ and no relaxation was performed.  The SWM \cite{PhysRev.108.612, PhysRev.109.272} tight-binding models were fit using a weighted least-squares fit on a uniform 24$\times$24 k mesh in a $\sim$3 eV window around the K-point, a fine mesh along the $\Gamma$-K-M path, which received a 30$\times$ weight, and an extremely fine mesh along the same path which received 100$\times$ weight in order to accurately reproduce the low energy band structure. 

\section{Results}

\subsection{Superconducting phase diagram}
We focus on electron-doped rhombohedral tetralayer graphene. Figure~\ref{fig:phase_diagram_1}
compares the intravalley instability of the spin- and valley-polarized
model with the intervalley instability of the valley-unpolarized
reference model. For the parameter set shown, the dielectric constant
is $\varepsilon_r=6$ and the pairing cutoff is $W=2.5~\mathrm{meV}$. The calculation was performed at fixed chemical potential and mapped to the density at $T=0$. 
As discussed above, the two calculations use different normal-state
valley configurations and therefore compare pairing tendencies rather
than establish a phase boundary between competing condensates of the
same normal state.

The two channels exhibit a clear contrast between transition temperature
and extent of the superconducting region. The maximum intravalley
transition temperature is $T_c^{\rm intra}=17.35~\mathrm{K}$, approximately
four times the maximum intervalley value,
$T_c^{\rm inter}=4.49~\mathrm{K}$. Nevertheless, the intervalley
instability extends over a substantially broader range of carrier
density and displacement field. This broader region is consistent with
the absence of an energy mismatch between time-reversed partners:
an attractive intervalley channel retains the logarithmic Cooper
enhancement, whereas intravalley pairing must overcome
trigonal-warping-induced pair breaking. The white regions in
Fig.~\ref{fig:phase_diagram_1} indicate points for which no superconducting transition is
resolved above our numerical threshold, $T_c > 0.05$ K.
For intravalley pairing, this suppression is primarily caused by the
energy mismatch generated by trigonal warping, which cuts off the
Cooper logarithm and can eliminate the weak-coupling instability
altogether; in parts of the phase diagram the corresponding pair Fermi
surface also disappears. For intervalley pairing, by contrast,
time-reversal symmetry ensures perfect particle--particle nesting and
there is no analogous pair-breaking mismatch. Thus, whenever an
attractive intervalley channel is present, a superconducting instability
exists in principle at arbitrarily low temperature, and the white
regions correspond to transition temperatures falling below our
numerical threshold rather than to the absence of the Cooper
instability.

The leading intravalley instability has predominantly $p_x-ip_y$
character, belonging to the $r=-1$ representation of $C_3$ in our
valley convention, consistent with previous calculations of
repulsion-driven intravalley pairing in rhombohedral
graphene~\cite{Yang2025FFLO,MayMann2026ValleyHelical}.
The preference for one chirality is associated with the
Berry-curvature-dependent complex phases of the Bloch form factors
in the valley-polarized band. Since the normal state already breaks
time-reversal symmetry, the two opposite chiralities are not required
to be degenerate. Their relative pairing strengths depend on the
interplay between these geometric phases and the momentum dependence
of the screened interaction; the preferred chirality is therefore
not determined by Berry curvature alone~\cite{MayMann2026PairingTopology}.

In contrast, the dominant intervalley instability over the main
superconducting region belongs to the two-dimensional $E$
representation of $C_{3v}$, consistent with earlier findings of
degenerate unconventional intervalley pairing channels in
rhombohedral graphene~\cite{Ghazaryan2023Multilayer,MayMann2026ValleyHelical}, although $A_1$ and $A_2$ irreps are found in the low density and high displacement field region. Writing the order parameter in the $E$ irrep as
$\Delta(\mathbf{k})=\eta_1 f_1(\mathbf{k})+\eta_2 f_2(\mathbf{k})$,
where $(f_1,f_2)$ form a basis of $E$, the linearized gap equation
determines the degenerate pairing subspace but not the relative
amplitudes or phase of $(\eta_1,\eta_2)$. Whether the superconducting
state selects a chiral combination $f_1\pm i f_2$ or a real nematic
combination requires an analysis of the quartic terms in the
Ginzburg--Landau free energy and remains beyond the present
linearized treatment. 

Recent quantum-oscillation measurements in rhombohedral tetralayer graphene
show that the superconducting state does not emerge from the simple
single-pocket circular quarter metal observed at higher carrier density.
Instead, superconductivity lies entirely within a lower-density ``multitone''
normal state characterized by a substantially more complex quantum-oscillation
spectrum~\cite{Kalantre2026Fermiology}.  This trend is
consistent with our intravalley results: the dominant superconducting
instability occurs predominantly in regions with annular or three-lobed Fermi
surfaces, whereas superconductivity is strongly suppressed in the
simply-connected single-pocket regime. While the experimentally observed
multitone state cannot yet be identified unambiguously with either of these
specific Fermi-surface topologies, both experiment and our calculation point
to Fermi-surface reconstruction away from the simple circular quarter metal
as an important ingredient for superconductivity.

The intravalley superconducting regions lie close to selected Lifshitz
boundaries of the pair dispersion, while the intervalley reference
exhibits a broader distribution around its ordinary-band Lifshitz
boundaries. To understand this contrast, we examine separately the
available pairing phase space, the energy mismatch, and the
symmetry-resolved interaction.

\subsection{Pair fermiology and trigonal-warping pair breaking}
\label{sec:pair_fermiology}

The connection between superconductivity and the reconstructed normal-state
fermiology can be sharpened by distinguishing the ordinary Fermi surface from
the Fermi surface relevant to intravalley Cooper pairing. As discussed in
Sec.~\ref{sec:trigonal_warping}, the latter is defined by the symmetrized
dispersion, $\varepsilon_+(\mathbf{k})=0$, and therefore need not coincide with the single-particle Fermi surface.
Consequently, the topology and density of states of the \emph{pair Fermi
surface} provide a more direct measure of the phase space available to the
intravalley instability.

Figure~\ref{fig:dos_plot} compares the ordinary single-particle density of states
$N_K$ with the pair density of states $N_{\rm pair}$ in the valley-polarized
model. The two quantities exhibit distinct structures and their associated
Lifshitz boundaries do not generally coincide. In particular, the pair
dispersion can undergo a topology change independently of the underlying
single-particle band, and in parts of the low-density, large-displacement-field
regime the pair Fermi surface disappears altogether even though an ordinary
Fermi surface remains present. In such regions the weak-coupling intravalley
Cooper instability is absent within the present pair-Fermi-surface
description.

\begin{figure}
    \centering
    \includegraphics[width=1.0\linewidth]{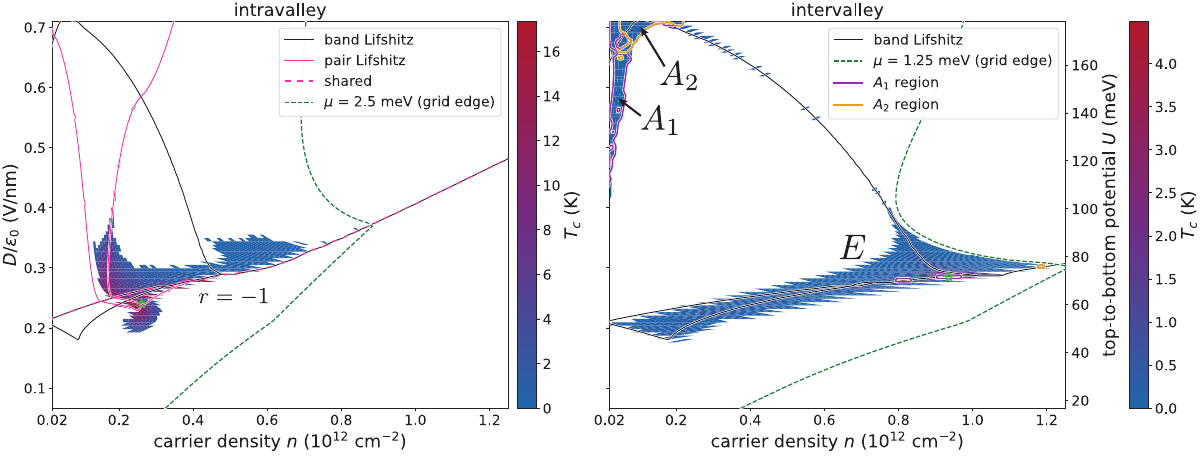}
    \caption{Phase diagrams of intravalley and intervalley pairing in four layer graphene, calculated with (upper) cutoff $W = 2.5$ meV and $\varepsilon_r=6$. On the left, the phase diagram for intravalley superconductivity in the polarized model $\gamma \neq 0$. Superconductivity is seen to occur close to the Lifshitz lines of the pair Fermi surface. On the right, intervalley pairing in the $\gamma = 0$ model is shown. The superconducting region extends over a larger parameter space due to the logarithmic instability of the un-pair-broken kernel. However, the critical temperature is much lower. The maximum $T_c$, marked with little green stars, are 17.35 K and 4.49 K for intravalley and intervalley respectively. The dominant pairing symmetries of the superconducting regions are marked. All calculated intravalley points were in the $r=-1$ irrep. The main intervalley dome was in the $E$ irrep and at low density and high field the $A_1$ and $A_2$ irreps are dominant. The calculation was performed at fixed $\mu$ and mapped at $T=0$. The boundary of the calculation is shown by the green dashed line. As the superconducting region with $T_c>0.05$ K does not intersect this line, we do not expect significant superconducting regions outside of it. }
    \label{fig:phase_diagram_1}
\end{figure}

\begin{figure}
    \centering
    \includegraphics[width=1.0\linewidth]{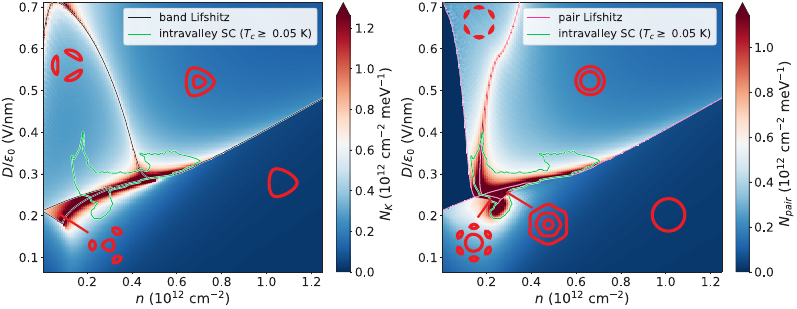}
    \caption{Density of states and pair density of states for the valley polarized model $\gamma\neq0$ at zero temperature. In the pair density of states, at low density but high field, the trigonal warping is strong enough that there are regions with no pair Fermi surface. Representative Fermi surfaces are shown in red for each region. The regular dispersion has regions with one pocket, three pocket, and four pockets, as well as an annular region. Not shown is a small sliver of a six pocket region at very low density and high displacement field in the top left of the figure. The pair dispersion has a different Fermiology, with one, six and seven pocket regions as well as annular and three nested pocket topologies. The intravalley superconducting solutions are shown outlined in green, and is seen to exist mostly in the annular and three pocket regions of the regular Fermi surface. Its shape is seen to correlate strongly with the density of states of the pair Fermi surface. }
    \label{fig:dos_plot}
\end{figure}

The superconducting regions in Fig.~\ref{fig:phase_diagram_1} correlate more
closely with the structure of $N_{\rm pair}$ than with the ordinary density
of states. Enhanced pair phase space occurs near several pair Lifshitz
boundaries, including the regions associated with annular and three-lobed
single-particle Fermi surfaces where the dominant intravalley superconducting
instability is found. This provides a natural connection between the
Fermi-surface reconstruction discussed above and the enhancement of
intravalley superconductivity. Importantly, however, a large $N_{\rm pair}$
is not by itself sufficient to produce superconductivity.

\begin{figure}
    \centering
    \includegraphics[width=0.7\linewidth]{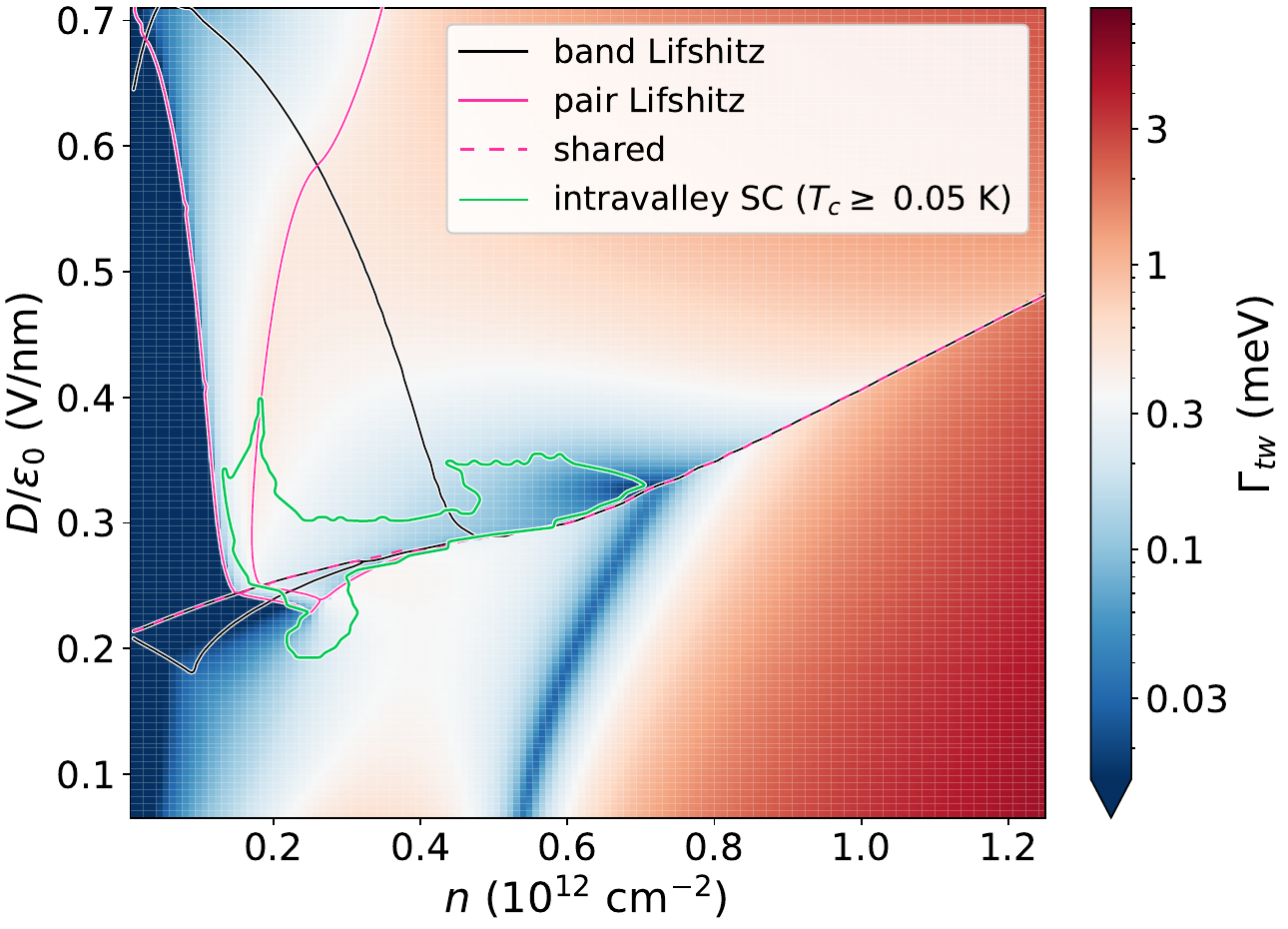}
    \caption{Pair breaking metric $\Gamma_{\textrm{tw}}$ for 4-layer electron doped graphene. The region with the least pair breaking but still with a pair Fermi surface is near the shared Lifshitz transition of the pair and band Fermi surfaces, as well as just below at lower densities. These regions coincide with those showing an intravalley pairing instability above. }
    \label{fig:gamma_tw}
\end{figure}

The second ingredient is the energy mismatch between the two members of an
intravalley Cooper pair, $\varepsilon_-(\mathbf{k})$, which originates predominantly from trigonal warping. Figure~\ref{fig:gamma_tw}
shows the corresponding pair-breaking scale $\Gamma_{\rm tw}$ across the same
density--displacement-field plane. The weakest pair breaking occurs close to
the shared single-particle and pair Lifshitz boundary and extends toward the
lower-density side of this boundary. These regions overlap closely with the
dominant intravalley superconducting regions in
Fig.~\ref{fig:phase_diagram_1}. By contrast, regions with substantial
$\Gamma_{\rm tw}$ generally show strongly reduced or unresolved $T_c$ even
when the pair density of states remains appreciable.

The calculated phase diagram therefore reflects the competition between two
effects generated by the reconstructed fermiology. Fermi-surface
reconstruction can enhance the available Cooper-pair phase space through a
large $N_{\rm pair}$ and favorable pair Lifshitz transitions, while the same
band structure also generates a momentum-dependent mismatch that cuts off the
Cooper logarithm. Intravalley superconductivity is strongest where these two
requirements are simultaneously satisfied. This explains why the dominant superconducting instability is concentrated
near selected reconstructed-Fermi-surface regimes rather than throughout the
entire high-density-of-states region, and why the ordinary Fermi-surface
topology alone is insufficient to determine the superconducting phase
boundaries.

\subsection{Symmetry-resolved pairing interactions}
\label{sec:pairing_interaction}

Having established how pair fermiology and trigonal-warping-induced mismatch
control where intravalley superconductivity can occur, we next examine the
interaction itself. The symmetry-resolved eigenvalues defined in
Sec.~\ref{sec:formfactor_interaction} isolate the structure of the projected
pairing interaction from the explicit pair-DOS weight and pair-breaking
kernel. They therefore provide a useful measure of which pairing channel is
favored once an instability becomes possible, and of how the
form factors dress the effective interaction in the different symmetry
channels.

\begin{figure}
    \centering
    \includegraphics[width=1.0\linewidth]{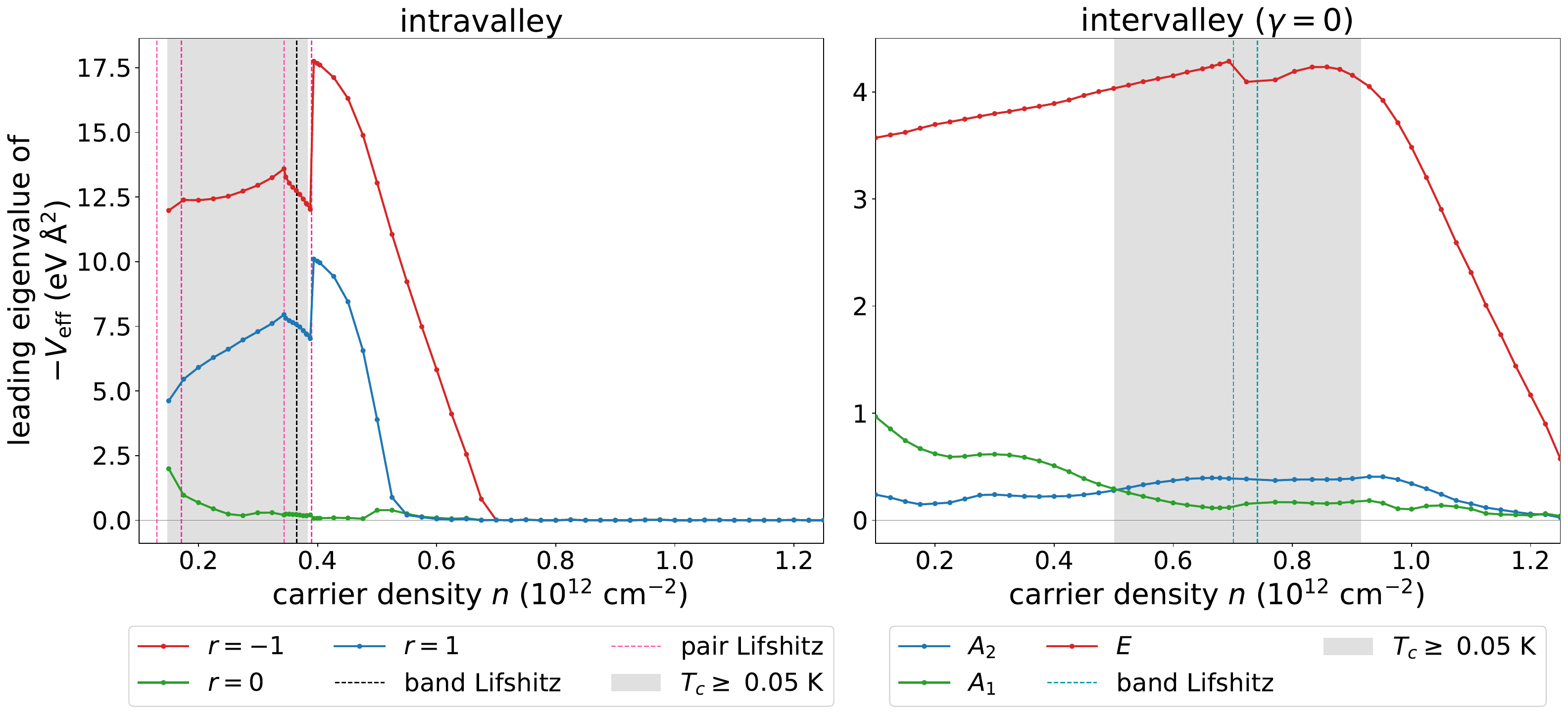}
    \caption{Leading attractive eigenvalues of the interaction restricted to the relevant Fermi surface at the displacement field $D/\varepsilon_0 = 0.2748$ V/nm. It is seen that the form factors at the Fermi surface cause a large increase in the pairing strength for low densities. In particular, the form factors in the intravalley case favor the $r=-1$ channel leading to $p_x - ip_y$ superconductivity as the leading instability. They are also much larger than the intervalley case, which has a leading instability in the $E$ irrep. Jumps in the effective interaction here are due to disappearances of one Fermi surface pocket at the Lifshitz transition.}
    \label{fig:interaction_channels}
\end{figure}

Figure~\ref{fig:interaction_channels} shows a representative density cut at
$D/\varepsilon_0=0.2748~\mathrm{V/nm}$. For intravalley pairing, the leading
attractive interaction at low density lies in the $r=-1$ sector, followed by
$r=+1$, while the $r=0$ channel is substantially weaker. In the
approximate angular-momentum language introduced above, the dominant
$r=-1$ sector corresponds primarily to a $p_x-ip_y$ state. The splitting
between the $r=\pm1$ channels is allowed because the valley-polarized normal
state already breaks time-reversal symmetry. Moreover, the intravalley
Cooper vertex retains the complex phase structure of the Bloch overlaps,
allowing the Berry-curvature-dependent form-factor phases to distinguish
the two chiralities.

The intervalley interaction exhibits a qualitatively different symmetry
structure. Its leading attractive channel belongs to the two-dimensional
$E$ representation of $C_{3v}$ over the main low-density region. In this
case, time-reversal symmetry relates the two valleys and makes the Cooper
form-factor product real and non-negative. The linearized calculation
therefore identifies the degenerate $E$ subspace but does not determine
how its two components combine below $T_c$; a distinction between chiral
and nematic intervalley states requires quartic terms in the free energy.

A second important feature of Fig.~\ref{fig:interaction_channels} is the
difference in the overall interaction scale. Over the low-density regime,
the leading intravalley attractive eigenvalue is substantially larger than
its intervalley counterpart, consistent with the significantly larger
maximum $T_c$ found for intravalley pairing in
Fig.~\ref{fig:phase_diagram_1}. The interaction eigenvalues also change
sharply across Lifshitz transitions, where Fermi-surface pockets appear or
disappear and the momentum-space region sampled by the projected
interaction changes abruptly. At larger density, the attractive
intravalley eigenvalues become strongly reduced as the relevant Fermi
surface moves away from the region where the Bloch form factors have their
strongest momentum dependence.

Importantly, the region with an appreciable attractive interaction does not
coincide exactly with the superconducting region. In particular, an
attractive intravalley channel can persist even where $T_c$ is strongly
suppressed or unresolved. This demonstrates that the interaction determines
the preferred pairing symmetry and its intrinsic strength, but does not by
itself determine the superconducting phase boundary. The latter results from
its combination with the available pair phase space and the
trigonal-warping-induced energy mismatch discussed above.

The pronounced difference between the intra- and intervalley projected
interactions originates from the momentum dependence and phase structure of
the Bloch wave functions. To further characterize the underlying band
geometry, we next examine the Berry curvature, quantum metric, and deviation
from the ideal Landau-level trace condition at representative points in the
phase diagram.

\subsection{Local quantum geometry}
\label{sec:local_geometry}

\begin{figure}
    \centering
    \includegraphics[width=0.9\linewidth]{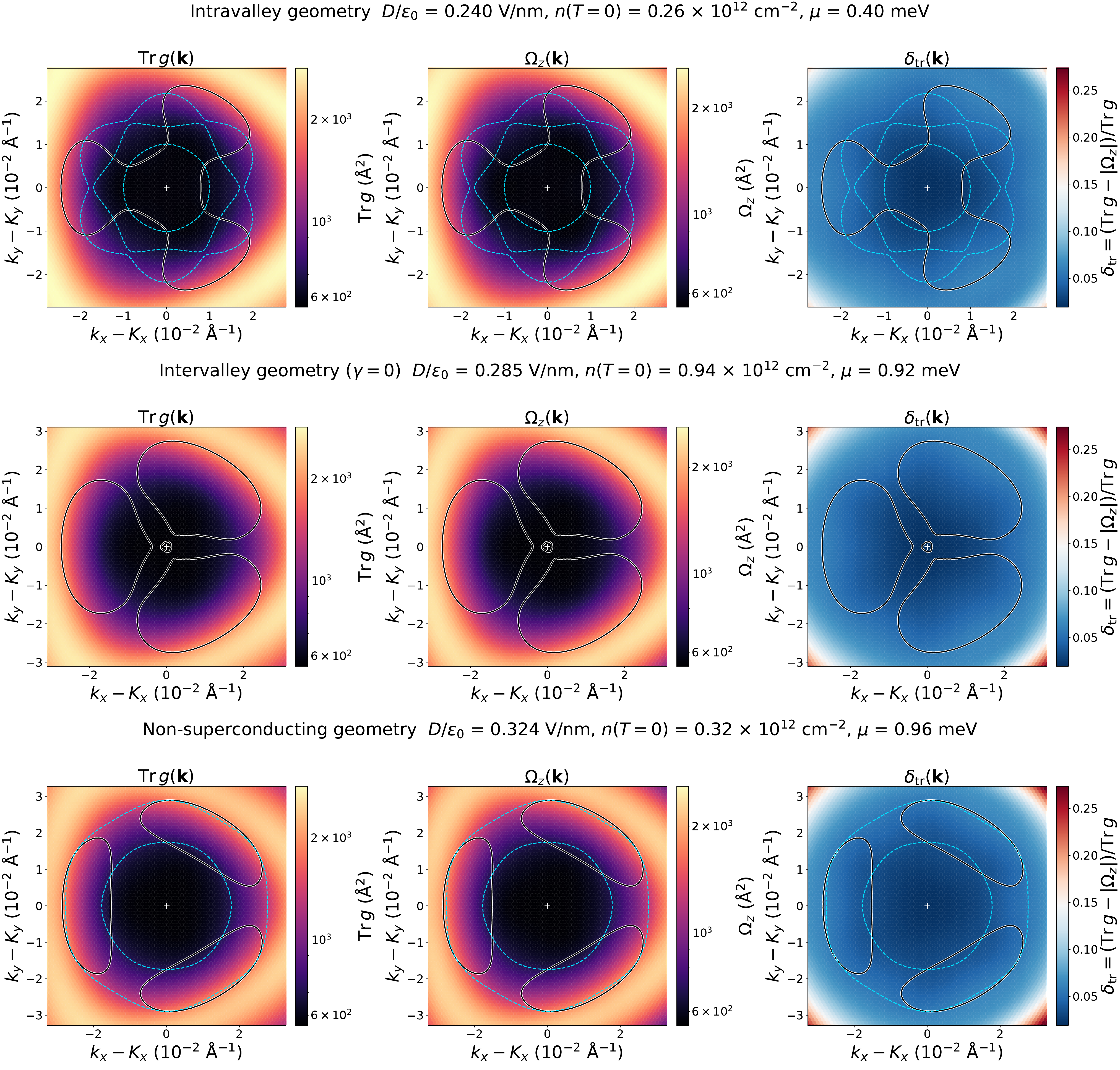}
    \caption{Trace of the quantum metric tensor, Berry curvature, and deviation from ideal geometry for the max $T_c$ points for intravalley and intervalley, along with a representative non superconducting point ($T_c < 10^{-3}K$). In the range in which we found superconductivity with $D/\varepsilon_0 > 0.15$ V/nm, the geometric quantities do not vary much with displacement field.}
    \label{fig:geometry}
\end{figure}

The large intravalley pairing interaction found above motivates a more direct
examination of the geometry of the underlying Bloch states. As discussed in
Sec.~\ref{sec:quantum_geometry}, the chiral model of rhombohedral graphene
provides a particularly simple reference point, where the low-energy band realizes the
same quantum geometry as the lowest Landau level. The realistic bands depart from this idealized limit through remote
hoppings, trigonal warping, and displacement-field-induced band
reconstruction. Figure~\ref{fig:geometry} shows
$\operatorname{Tr}g(\mathbf{k})$, $\Omega(\mathbf{k})$, and the normalized
trace-condition violation $\delta_{\rm tr}(\mathbf{k})$ for three representative points: the maximum-$T_c$ intravalley point, the
maximum-$T_c$ intervalley point, and a nonsuperconducting point.

At the superconducting points, the Berry curvature and quantum metric retain
the same qualitative momentum-space structure and are relatively smooth over
the momentum region sampled by the relevant Fermi surfaces. The trace
condition remains approximately satisfied, with
$\delta_{\rm tr}$ typically of order ten percent over much of this region.
The realistic bands therefore remain close to the ideal Landau-level
geometric limit, even though both $\Omega(\mathbf{k})$ and
$g_{ij}(\mathbf{k})$ acquire finite momentum dependence.

The comparison with the nonsuperconducting point is particularly useful.
Its local Berry curvature, quantum metric, and trace-condition violation are
broadly similar to those at the superconducting points. Thus, approximate
Landau-level-like local geometry is not, by itself, sufficient to determine
whether superconductivity occurs. Rather, it should be viewed as providing a
favorable geometric background that strongly dresses the projected
interaction, while the actual superconducting instability is additionally
controlled by pair fermiology and trigonal-warping-induced pair breaking.

Taken together, the results support a picture in which the low-energy bands
of rhombohedral graphene retain an approximately Landau-level-like quantum
geometry over the relevant momentum range, consistent with the simplified
chiral model. This geometry enhances and differentiates the projected pairing
channels, but the detailed superconducting phase boundaries emerge only after
combining the form-factor-dressed interaction with the available pair phase
space and the momentum-dependent energy mismatch.

\section{Discussion and conclusions}
\label{sec:discussion}

Our results distinguish three ingredients controlling superconductivity in
rhombohedral graphene: the form-factor-dressed interaction, the phase space
of the symmetrized pair dispersion, and the mismatch between the paired
states. Within the models considered here, the polarized intravalley
$p_x-ip_y$ channel has a stronger attractive interaction and a higher maximum
$T_c$ than the intervalley reference state, but occupies a narrower region
of the density--displacement-field phase diagram. This contrast reflects
the different pairing kinematics. Intervalley pairing between
time-reversed partners retains the Cooper logarithm whenever an attractive
channel is present, whereas intravalley pairing must overcome the mismatch
generated by trigonal warping. Its strongest instabilities therefore occur
where favorable pair phase space near selected Lifshitz boundaries
coexists with sufficiently weak pair breaking. A large attractive
interaction or an enhanced single-particle DOS alone is consequently
insufficient to determine the superconducting phase boundaries.

This picture provides a useful perspective on the experimental
fermiology. The predominantly chiral intravalley instability is compatible
with the reported signatures of time-reversal-breaking superconductivity,
although these signatures do not uniquely identify the pairing
mechanism~\cite{Han2025ChiralSC}. For the tetralayer parameters studied
here, the dominant instability occurs in annular or three-lobed regimes
and is strongly suppressed over much of the simply connected regime.
This resembles the experimental finding that superconductivity lies
within a lower-density multitone state rather than the circular quarter
metal observed at higher density~\cite{Kalantre2026Fermiology}.
The agreement is qualitative, however: the measured multitone spectrum
is not reproduced by the simple polarized annular, three-pocket, or
nematic states examined in that work. Moreover, pentalayer experiments
find superconducting regions on both sides of a circular-to-annular
Lifshitz transition, accompanied by a strong effective-mass enhancement
near the boundary~\cite{Zheng2026Lifshitz}. Thus, suppression of
superconductivity in the simply connected regime is not universal.
Taken together, these observations motivate the distinction between
ordinary Fermi-surface topology and the combined interaction, pair
phase space, and mismatch that determine the superconducting instability.

The quantum-geometric results further clarify this distinction.
At the representative points studied, the realistic bands approximately
satisfy the ideal trace condition, consistent with the Landau-level-like
geometry obtained in simplified descriptions of rhombohedral
graphene~\cite{bernevig2025berrytrashcanmodelinteracting,Jiang2026IdealQuantumGeometry}.
This connects our calculation to model studies showing that
Landau-level-like Bloch form factors can enhance unconventional
Kohn--Luttinger pairing~\cite{Jahin2026Enhanced}. Nevertheless, comparable
local geometry also occurs at the nonsuperconducting reference point.
Approximate trace-condition saturation therefore does not identify the
superconducting region, nor does it establish uniform geometry or a
Landau-level form factor at all momentum transfers. The screened Cooper
interaction depends on the full finite-momentum overlaps, including their
complex phases, rather than solely on the local Berry curvature and
quantum metric. These results support a distinction between geometric
properties that can favor attractive chiral channels and the pairing
kinematics that determine whether this attraction produces a
superconducting instability.

We have not optimized finite displacements of the intravalley pairing
center away from $\mathbf K$. Trigonal warping can favor an additional
pair momentum $\mathbf q$ that partially compensates the energy mismatch,
producing an incommensurate superconducting state with total crystal
momentum $2\mathbf K+\mathbf q$, modulo a reciprocal lattice
vector~\cite{Yang2025FFLO}. Here $2\mathbf K$ is the momentum inherent
to pairing within a single valley, while $\mathbf q/2$ represents the
displacement of the pairing center. Although $|\mathbf q|\ll|2\mathbf K|$,
this additional shift can affect the low-energy pair mismatch and hence
the transition temperature and superconducting boundaries.

Finally, valley polarization is imposed phenomenologically, so
interaction-induced reconstruction of the normal-state bands is not
determined self-consistently. Accordingly, the intra- and intervalley
calculations compare instabilities of different normal states rather
than their thermodynamic competition. For the intravalley calculation,
we assume that the valley-polarization energy scale is much larger
than the superconducting pairing scale, allowing polarization to be
treated as a fixed property of the normal state. When these scales
become comparable, the feedback between superconductivity and valley
order requires a coupled, self-consistent treatment. Partial valley
polarization generally lifts the degeneracy between opposite-valley
pairing partners and removes the symmetry-protected particle--particle
nesting of the intervalley channel. This mismatch is expected to
suppress intervalley pairing relative to intravalley pairing. Valley polarization is nevertheless not a prerequisite for intravalley
superconductivity. Recent theory has shown that repulsive interactions
and quantum geometry can favor intravalley over intervalley pairing
even in a time-reversal-symmetric, valley-unpolarized normal
state~\cite{MayMann2026ValleyHelical}. In this case, intravalley pairing
develops in both $K$ and $K'=-K$ valleys with opposite chiralities,
producing a valley-helical superconducting state. 

\begin{acknowledgements}
  The authors thanks Ammar Jahin for helpful discussions. Y.Z. was supported by the Max Planck partner lab for quantum materials. The work at LANL was carried out under the auspices of the U.S. DOE NNSA under contract No. 89233218CNA000001 through the LDRD Program, and was supported by the Center for Nonlinear Studies at LANL, and was performed, in part, at the Center for Integrated Nanotechnologies, an Office of Science User Facility operated for the U.S. DOE Office of Science, under user proposals $\#2018BU0010$ and $\#2018BU0083$.

  Generative AI (Claude Fable 5.1, Opus 5.5) was used to assist in the programming. The authors claim sole responsibility for the accuracy of the implementation. It was also used to check the manuscript for errors in the writing. 

\end{acknowledgements}

\bibliography{refs.bib}

\end{document}